\documentstyle[preprint,aps]{revtex}
\begin{document}

\draft
\title{
Astrophysical Reaction Rates for
$^{10}$B(p,$\alpha$)$^{7}$Be and $^{11}$B(p,$\alpha$)$^{8}$Be
From a Direct Model
}

\author{T. Rauscher}
\address{
Institut f\"ur Kernchemie, Universit\"at Mainz, 
Germany\\
and\\
Institut f\"ur theoretische Physik, Universit\"at Basel, Switzerland 
\footnote{current address}}

\author{G. Raimann}
\address{
Department of Physics, The Ohio State University, Columbus, OH, USA
}

\maketitle

\begin{abstract}
The reactions $^{10}$B(p,$\alpha$)$^{7}$Be and
$^{11}$B(p,$\alpha$)$^{8}$Be are studied at thermonuclear energies
using DWBA calculations. For both reactions, transitions to the ground
states and first excited
states are investigated. In the case of $^{10}$B(p,$\alpha$)$^{7}$Be, a
resonance at $E_{Res}=10$ keV can be consistently described in the
potential model, thereby allowing the extension of the astrophysical 
$S$-factor data to very low energies. 
Strong interference with a resonance at about $E_{Res}=550$ keV require
a Breit-Wigner description of that resonance and the introduction of
an interference term for the reaction $^{10}$B(p,$\alpha_1$)$^{7}$Be$^*$.
Two isospin $T=1$ resonances (at $E_{Res1}=149$ keV
and $E_{Res2}=619$ keV) observed in the $^{11}$B+p reactions necessitate
Breit-Wigner resonance  and interference terms 
to fit the data of the $^{11}$B(p,$\alpha$)$^{8}$Be reaction. 
$S$-factors and thermonuclear reaction rates are given for each reaction.
The present calculation is the first consistent parametrization for the
transition to the ground states and first excited states at low energies.
\end{abstract}

\pacs{25.40.Hs, 24.10.Eq, 24.50.+g, 28.52.Cx}

\section{Introduction}

The determination of the astrophysical $S$-factor
of the reaction $^{10}$B(p,$\alpha$)$^7$Be at thermonuclear energies
is important in
several respects. In the search for advanced fusion reaction fuels the 
reaction
$^{11}$B + p $\rightarrow$ $^8$Be + $\alpha$ $\rightarrow$ 3$\alpha$
is discussed as a promising candidate for a relatively
clean fusion fuel~\cite{rae81,fel88}. However, natural boron contains 19.7\%
$^{10}$B which produces $^7$Be contaminations via the
$^{10}$B(p,$\alpha$)$^7$Be reaction. Therefore, for a full understanding
of the feasibility of boron as a fusion fuel one has to 
consider the rate for the latter reaction as well~\cite{pet75}.

The importance of the
$^{10}$B(p,$\alpha$)$^7$Be reaction in astrophysical scenarios results from the
fact that it is the dominant process for 
the destruction of $^{10}$B. It
was also claimed that it could be usefully
incorporated in explaining the abundances of boron isotopes
including the present theory of spallative generation of $l$ 
elements~\cite{you91,ree71}.
Furthermore, in
theoretical investigations of primordial abundances of elements, its rate
has to be incorporated in the reaction networks employed in
nucleosynthesis calculations for inhomogeneous big
bang scenarios~\cite{thi91,app93}.

Experimental results for the $^{10}$B(p,$\alpha$)$^7$Be reaction at
energies below 1 MeV are scarce. There have been measurements of the
$^{10}$B(p,$\alpha_0$)$^7$Be cross sections with
sometimes inconsistent results in the energy ranges 220 keV $< E_p <$ 480 
keV~\cite{bur50}, 60 keV $< E_p <$ 180 keV~\cite{sza72}, 70 keV $< E_p <$ 205
keV~\cite{bac55}, and, more recently, 120 keV $< E_p <$ 480 keV~\cite{you91}.
However, these measurements do not extend very far into the region dominated
by the $J^{\pi}$=5/2$^+$ resonance ($E_x = 8.701$ MeV in
$^{11}$C~\cite{wie83}) which is of great importance for astrophysics.
In our calculations we therefore focused on
the description of the resonance and the reproduction of most recent
data~\cite{kna92a,kna92b,ang93} at very low energies. The only available
data measuring the $\alpha _1$ contribution to the $S$-factor are given
in Ref.~\cite{ang93b}.

The reaction $^{11}$B(p,$\alpha$)$^8$Be has been measured below 1 MeV
\cite{dav79}, \cite{beck87}, and most recently \cite{ang93}.
It should be noted that the effect of electron screening~\cite{ang93}
increases the very low energy cross sections considerably and thus
has a major impact on a reaction's significance as a terrestrial fusion
fuel.

\section{Method}

It is commonly accepted that nuclear reactions for energies above about
20 MeV mainly proceed via a direct mechanism. For intermediate energies,
however, distinct levels of the compound system are populated, resulting
in many cases in pronounced resonances in the excitation functions. 
For astrophysically relevant energies, typically 
sub-Coulomb energies of a few keV or tens of keV, compound mechanisms
are often very important. However, direct transitions can
also be important at stellar energies. For example, the reactions of the 
pp-chains in the Sun are known to be mainly dominated by such direct
mechanisms~\cite{tom67,bah89}. Recent theoretical investigations of a number
of sub-Coulomb transfer reactions~\cite{ohu89,rai90,her91,ohu91,rau92}
have also shown that they can be described by a direct reaction
potential model like the distorted wave Born approximation (DWBA).
Although ``direct reaction'' is often used synonymously with 
``non-resonant'',
the same potentials that describe the direct mechanism can give rise 
to resonances corresponding to energy levels in the projectile-target 
system. Such potential resonances
are very broad at energies above the Coulomb barrier and do not alter
cross sections significantly within several keV (or even a few MeV). 
However, due to the
Coulomb barrier they become small at low energies, with typical widths
of a few keV. 
A DWBA description of a resonance
structure at thermonuclear energies of a three-nucleon transfer reaction
is given in~\cite{rau92}. In the sub-Coulomb energy range of 
(p,$\alpha$) reactions the DWBA method has previously only been used to
analyze non-resonant parts of the excitation 
functions~\cite{rai90,her91}. We want to emphasize, though, that it is 
possible 
to reproduce resonance features in a DWBA calculation.

For the results in this paper we utilized the zero-range DWBA code
TETRA~\cite{tetra}.
The differential cross section
for the transfer reaction
$a+A \rightarrow b+B$ with $a-x=b$, $A+x=B$ (stripping) using light
projectiles and ejectiles (for $a \leq 4$ and $x=1$ or $x=3$) is given
in
zero-range DWBA by~\cite{sat83,gle83}
\begin{equation}
{d\sigma \over d\Omega} = {\mu_{\alpha}\mu_{\beta} \over
(2\pi\hbar^{2})^{2}}{k_{\beta} \over k_{\alpha}}
{2J_{B}+1 \over 2J_{A}+1} \sum_{\ell sj}C^{2}{\cal S}_{\ell j}N
{\sigma_{\ell sj}(\vartheta) \over 2s+1}
\end{equation}
with the usual zero-range normalization constant 
$N = 1/2 \; aD_{0}^2 = 5.12 \cdot 10^5$~\cite{hoy85}. 
The spectroscopic factors ${\cal S}_{\ell j}$
relevant for our calculations were taken from~\cite{kur75}.

Important for the success of our potential model is the fact that
the input data for the optical potentials can be taken from realistic
models, i.e. from semimicroscopic
or microscopic formalisms such as the folding-potential model,
the Resonating Group Method (RGM), or the Generator Coordinate Method
(GCM).
In this respect the potential model
combines the first-principle approach of a microscopic theory with the
flexibility of a phenomenological method.

In this work we used the method of the folding potential~\cite{ohu91,sat79}
to obtain
the optical potentials in the entrance and exit channel as well as the
potential for the bound state.
The folding potential is given by~\cite{ohu91}
\begin{equation}
U_{F}({\bf r})=\lambda\int d{\bf r}_{A}\int d{\bf r}_{a}\;
\rho_{A}({\bf r}_{A})\;\rho_{a}({\bf r}_{a})\;
V(E,\rho_{A},\rho_{a},s=|{\bf r}+{\bf r}_{a}-{\bf r}_{A}|)
\quad.
\end{equation}
In this expression {\bf r} is the separation of the centers of mass of
the two
nuclei in the channel, $\rho_{a}$ and $\rho_{A}$
are the respective nucleon densities and $\lambda$ is the adjustable
strength factor. The factor $\lambda$ differs slightly from the value of 
1 because it accounts for the
effects of antisymmetrization and the Pauli principle.
The effective NN interaction $V$ for the
folding procedure was of the DDM3Y type~\cite{kob84} and the density
distributions
were taken from Ref.~\cite{dev87} unless noted otherwise.
For the bound state potentials $\lambda$ is fixed, since the known binding
energy of the transferred particle $x$, a triton, in $^{10}$B or $^{11}$B, 
respectively, has to be reproduced. 

While the resonant cross section (or $S$-factor) of 
$^{10}$B(p,$\alpha_0$)$^7$Be
can be reproduced well by the DWBA alone, an interfering resonance 
in $^{10}$B(p,$\alpha_1$)$^7$Be$^*$ and -- due to their
unnatural parity --
the resonances in
$^{11}$B(p,$\alpha$)$^8$Be have to be treated explicitly assuming, e.g.,
single-level Breit-Wigner expressions (see Sect.\ III.B).

\section{Results for $^{10}$B(p,$\alpha$)$^7$Be}

\subsection{The Reaction $^{10}$B(p,$\alpha_0$)$^7$Be}

At high energies the transition to the ground state of $^7$Be 
($Q=1.15$ MeV~\cite{ajz90}, binding energy $E_{bind}=18.67$ 
MeV~\cite{wap85} of the triton in the $^{10}$B nucleus) was previously 
analyzed by means of the DWBA method~\cite{hau86}. For the transition to the
first excited state at sub-Coulomb energies, there has only
been a simplified DWBA calculation of the direct reaction contribution to 
the cross section which considered only the Coulomb
potential~\cite{web89}. 
To our knowledge, no calculation of the contribution of the
transition to the ground state at sub-Coulomb energies has been 
performed so far.

The spectroscopic factors ${\cal
S }_{\ell j}$ are listed in Table~\ref{tab1}.
For the calculation of the
folding potential in the $\alpha$-$^7$Be channel as well as for
the bound state (t-$^7$Be) the density distribution of $^7$Li~\cite{dev87}
was used instead of the unknown distribution of $^7$Be.

In our calculation, the strength factor $\lambda$ for the 
optical potential in the
proton channel was adjusted in such a way that the phase shift of the
optical s-wave at the resonance energy $E_{Res}=0.01$
MeV was $\pi/2$, thereby
producing the s-wave resonance previously suggested~\cite{you91}. The value 
of $\lambda$ in the alpha channel remained an open parameter 
since there are no elastic scattering data available for the
unstable nucleus $^7$Be.
The imaginary part of the potential in the proton channel is very
small because there are no other channels open at these low energies besides
the (p,$\gamma$) and (p,$\alpha$) channels. 
The complete set of optical
parameters is given in Table~\ref{tab2}.

For the transition to the first excited state of $^7$Be$^*$ ($Q=0.72$
MeV) the optical potential in the proton
channel stays the same as for the ground state transition. We also
assume that the potential for $^7$Be$^*+\alpha _1$ can, to first order, be
approximated by the potential for $^7$Be$+\alpha_0$. 

For $^{10}$B(p,$\alpha_0$)$^7$Be the resulting $S$-factors in the energy
range $E_{c.m.} \leq 150$ keV are shown in
Figs.~\ref{fig1a} and \ref{fig1b}. The experimental results~\cite{ang93}
are well reproduced with the
resonance width being about 15 keV, in agreement with other
measurements~\cite{wie83,ajz90}. It should be noted that the calculated
energy dependence of the $S$-factor is only valid for bare nuclei. Due
to electron screening, measurements to even lower energies would not show
the same Breit-Wigner like shape. The deviation of the experimental
data at the two lowest measured energies in Fig.~\ref{fig1a} is already
due to screening effects~\cite{ang93}.
The calculated differential cross sections
show the same isotropic behaviour as can be seen in the data of
Ref.~\cite{kna92b}.

\subsection{The Reaction $^{10}$B(p,$\alpha_1$)$^7$Be$^*$}

Using the same approach as described above we also calculated the
cross sections for the transition to the first excited state at 
energies $E_{c.m.} \leq 150$ keV. 
Interference effects with other resonances (especially a broad 5/2$^+$ level at
about 550 keV) were reported in~\cite{wie83} at slightly higher projectile
energies. In order to successfully reproduce the experimental data~\cite{ang93b}
while keeping unchanged all of the parameters entering the DWBA computation,
we had to include that 5/2$^+$ level in our calculation. This was
achieved by a single-level Breit-Wigner fit to the level at 550 keV and
by finally calculating a total cross section (or $S$-factor, respectively)
from the interference of the Breit-Wigner and the DWBA contributions.

The single level Breit-Wigner formula is given by~\cite{rol88}
\begin{equation}
\label{BW}
\sigma (E) = \pi \overlay{\mathchar'26}{\lambda}^2
{ 2 J_R + 1 \over (2 J_p + 1) (2 J_B + 1) }
   \times { \Gamma_p (E)_l \, \Gamma_\alpha \over
           (E - E_R)^2 + (\Gamma_{tot} (E) / 2)^2 } \quad .
\end{equation}
The quantities $J_R$, $J_p$ and $J_B$ denote the total angular momentum
of the resonance, of the incoming proton and of the $^{10}$B target nucleus,
respectively, $\Gamma_p (E)_l$ is the energy dependent proton partial width
of the resonance with orbital angular momentum $l$, $\Gamma_\alpha$ is the
energy independent alpha partial width, and $\Gamma_{tot} (E)$ is the
total width of the resonance as given by
\begin{equation}
\Gamma_{tot} (E) = \Gamma_p (E)_l + \Gamma_\alpha \quad .
\end{equation}
$\Gamma_\gamma$ would be small and is neglected here.
Due to the positive $Q$-value it is sufficient to consider the energy dependence
of only the proton partial width $\Gamma_p$. This energy dependence
can be described by~\cite{rol88}:
\begin{equation}
\Gamma_p (E)_l = { 2 \hbar \over R_n }
                 \left( { 2 E \over \mu} \right)^{1/2}
                 P_l (E, R_n) \theta_l^2 \quad ,
\end{equation}
with the penetrability
\begin{equation}
P_l (E,R_n) = { 1 \over F (E,R_n)^2 + G (E,R_n)^2 }
\end{equation}
given in terms of the regular and irregular Coulomb wave functions $F$ and $G$
and the nuclear radius $R_n$. For simplicity, $R_n$ will be derived here from
the Coulomb charge radius (light ion convention):
\begin{equation}
R_n = r_c  A_B^{1/3} \quad .
\end{equation}
The dimensionless quantity $\theta_l$ is the reduced proton width.

The available experimental data~\cite{wie83,ang93b} were not sufficient 
to yield an unambiguous fit within our calculation. 
A small alpha width leads to a larger proton
width and vice versa. For example, with a resonance energy of $E_{Res}=560$ keV
we obtain the following pairs of alpha partial width and proton reduced
width: $\Gamma_{\alpha}$=0.1 keV, $\theta_l$=0.49 and 
$\Gamma_{\alpha}$=500.0 keV, $\theta_l$=0.01 .
We were not able to distinguish between the 
different sets in our chi square fit. (However, the resonant (p,$\gamma$)
cross section~\cite{wie83} seems to favor a large reduced width for the
proton channel if the gamma width is assumed to be small).

Finally, we can write the energy-dependent $S$-factor as
\begin{equation}
S_{tot,\alpha_1} (E) = S_{DWBA} (E) + S_{BW} (E)
           - 2 \big[ S_{DWBA} (E) S_{BW} (E) \big]^{1/2}
           \cos \delta \quad .
\end{equation}
The phase shift $\delta$ is given by~\cite{rol74}
\begin{equation}
\label{delta}
\delta = \arctan \left( { 2 ( E - E_R ) \over \Gamma_{tot} (E) } \right)
                 - { \pi \over 2} \quad .
\end{equation}

The total $S$-factors and the DWBA and Breit-Wigner contributions
are shown in Fig.~\ref{fig2}. 
Although the experimental data~\cite{ang93b} are quite well reproduced at
higher
energies, the agreement slightly worsens toward lower energies. The experiment
seems to give a larger value for the width of the 10 keV resonance. 
This is caused by the
assumption that the imaginary part of the optical potential in the proton
channel is the same as for the transition to the ground state of $^7$Be.
Actually, the imaginary part for the $\alpha _1$ transition should be
slightly larger because more flux is going into the $\alpha _0$ channel
than into the relatively small
$\alpha _1$ channel which is included in the imaginary part used
for the ground state transition.
With a larger imaginary part the resonance
width is increased. This result is very sensitive to the depth of the
imaginary optical potential; with an increase by only 50 keV the resonance
structure is already flattened out completely.
However, in order to get an
upper limit on the contribution of this transition to the total $S$-factor
we used the same optical potential as for the ground state transition.

The ther\-mo\-nu\-cle\-ar reaction rate 
$N_A \left< \sigma v \right>$ is given in Table~\ref{tab6}, where $N_A$ is 
the Avogadro constant and the bracketed
quantity is the velocity averaged product of the cross section and the
relative
velocity of the interacting particles~\cite{rol88}. 
In Fig.~\ref{fig3} the ratio of the resulting rate at low temperatures to the
rate given in Ref.~\cite{cau75} is shown. This rate remained unchanged in a 
more
recent compilation~\cite{cau88} of reaction rates. Since the
$E_{Res}=0.01$ MeV resonance was not taken into account in
\cite{cau75}, its rates differ considerably from our new values in the 
corresponding temperature region. 
Our calculation shows that the contribution of the reaction 
$^{10}$B(p,$\alpha_1$)$^7$Be$^*$ is less than 10$^{-3}$ of the
rate for the ground state transition, and that the compiled rates have to be
revised. 

\section{The reactions $^{11}$B(p,$\alpha_0$)$^8$Be and
$^{11}$B(p,$\alpha_1$)$^8$Be$^*$}

\subsection{Available data}

Cross sections for this reaction were measured in Ref.~\cite{beck87}
($\alpha_0$ as well as $\alpha_1$) and recently 
to even lower energies in Ref.~\cite{ang93} (sum of
$\alpha_0$ and $\alpha_1$). In the latter case, the $\alpha_0$ $S$-factor 
amounted to only about
1\% of the $\alpha_1$ value at energies below 500 keV.
The reaction can proceed either via a direct 3$\alpha$-breakup, or via
a sequential decay involving the states of $^8$Be. It was demonstrated
in~\cite{beck87} that the direct 3$\alpha$-breakup makes no significant
contribution (less than 5 \%) to the total cross section at all the energies
investigated ($E_{c.m.} = 22$ to 1100 keV). It is therefore justified
to describe the reaction in terms of a quasi-stable $^8$Be nucleus 
interacting
with an $\alpha$-particle.

The $S$-factor of $^{11}$B(p,$\alpha_1$)$^8$Be$^*$ is dominated by
two $T=1$ resonances at $E_{c.m.} = 149$ keV ($J^\pi = 2^+$, orbital
angular momentum $l_R=1$) and
at $E_{c.m.} = 619$ keV ($J^\pi = 2^-$, $l_R=0$)~\cite{ajz90} in the $^{12}$C
compound system. 
Due to its unnatural parity,
the $E_{c.m.} = 619$ keV resonance would not be expected to
contribute to the $S$-factor of
the $^{11}$B(p,$\alpha_0$)$^8$Be reaction. 
In Ref.\ \cite{ajz90} the two resonances are quoted as pure $T=1$. Without
at least a small admixture of $T=0$, however, these resonances
could not decay into the $^8$Be + $\alpha$ channel. The concept of isospin
mixing has been thoroughly investigated and understood for the 1$^+$ states; 
experiments with pion scattering \cite{jak90,cot87} show that several excited 
$^{12}$C states exhibit quite considerable mixing. The two states in
question would only be weakly excited in pion scattering and have not been 
observed, but it is plausible that all high-lying $^{12}$C states are 
isospin mixed \cite{cle93}. 

Within the simple approach adopted here, the DWBA calculation ($T=0$) gives
the non-resonant contribution. The mostly $T=1$ resonances are reproduced
by single-level Breit-Wigner terms, and the $T=0$ admixture of the 619 keV
resonance ($l_R=0$) interferes with the mostly $l=0$ direct part. The
isospin mixing is accounted for implicitly in the Breit-Wigner approach
via the decay width into the $\alpha$-channel.

Lacking an appropriate representation for the direct
contribution, Ref.~\cite{beck87}
gives only a polynomial fit of the $S$-factor. In Ref.~\cite{ang93}, the
direct, non-resonant contribution is assumed to be energy independent, and 
the same formalism of a Breit-Wigner term plus an interference term is 
employed to describe the cross section. In this work,
the determination of the direct contribution is based on a more basic
calculation and the transition to the
ground state is described with the same set of parameters. 

\subsection{The Calculation}

The $Q$-value of this process is $Q = 8.59$ MeV~\cite{ajz90}, and the
binding energy is $E_{bind}=11.22$ MeV~\cite{wap85}; the spectroscopic
factors ${\cal S}_{\ell j}$~\cite{kur75} including all constants are
listed in Table~\ref{tab3}.

The calculation was performed using a $^8$Be density distribution
that was chosen \cite{sch94} so that folding it
with the triton density 
would reproduce the $^{11}$B distribution. 
The strength parameter $\lambda_\alpha = 1.21$ is fairly close to the 
preliminary result obtained for the triple-alpha reaction~\cite{sch94}. 

The strength factor $\lambda_p$ in the proton channel was kept
as a free parameter, since there are no low energy elastic scattering
data available suited for an optical potential fit.
Differential cross sections measured at one angle \cite{beck87}
confirm that elastic scattering below 400 keV in the center-of-mass is 
consistent
with the Rutherford scattering law. 

Due to the low energy in the proton channel both the imaginary
potential and a real spin-orbit term can be neglected. 
In the alpha channel, a non-zero
Saxon-Woods imaginary part was used (the geometry parameters were
roughly averaged between those for $\alpha$-$^7$Li and $\alpha$-$^9$Be, 
both taken from Ref.~\cite{per76}). 
The complete set of optical parameters is given in Table~\ref{tab4}.
In first order, the optical potentials are assumed to be identical for
the reactions to the $^8$Be ground state and to the first excited state.

Similarly to the approach for $^{10}$B(p,$\alpha_1$)$^7$Be$^*$, a
sum of Breit-Wigner terms and an interference term was used.
The total $S$-factor for the reaction to the first excited state of 
$^8$Be$^*$ consists of the contributions by both resonances 
($S_{Res1}(E)$ and $S_{Res2}(E)$ at $E_{c.m.} = 149$ keV and 
$E_{c.m.} = 619$ keV, with orbital angular momenta $l_R=1$ and $l_R=0$,
respectively), the non-resonant contribution $S_{NR}$ as calculated
by the DWBA, and an interference term \cite{ang93,rol74}:
\begin{equation}
S_{tot,\alpha_1} (E) = S_{Res1} (E) + S_{Res2} (E) + S_{NR} (E)
           - 2 \big[ S_{NR} (E) S_{Res2} (E) \big]^{1/2}
	   \cos \delta \quad .
\end{equation}
Each resonance is described by a single-level Breit-Wigner term
(Eq.~\ref{BW}) with a
fixed $\alpha$-partial width $\Gamma_\alpha$  and an energy dependent
proton partial width $\Gamma_p(E)$ expressed in terms of the penetrability
$P_l(E)$ and the reduced proton width $\theta_p$. 
The phase shift $\delta$ is given by Eq.~\ref{delta}.
The interference term is between the non-resonant component
and 
the $T=0$ fraction of the 619 keV resonance which is determined by the
Breit-Wigner fit. 
There is no interference with the 149 keV resonance since it has $l_R=1$.

The total $S$-factor for the reaction to the $^8$Be ground state
consists of only the $S$-factor contributions of the lower-energy resonance 
($l_R=1$) and the non-resonant term as calculated by the DWBA.

For a value of $\lambda_p$ almost identical to the one used for 
$^{10}$B(p,$\alpha$)$^7$Be, the
Breit-Wigner terms (including the energy dependent proton partial width)
were fitted to the experimental data of Refs.\
\cite{ang93} and \cite{beck87}.
The values of $\theta_p$, $\Gamma_\alpha$ and $E_{Res2}$ resulting from the 
fit are listed in Table~\ref{tab5}.

\subsection{Results}

The resulting $S$-factor curves are shown in Fig.~\ref{fig4b}.
The dashed lines
represent the non-resonant contribution of the DWBA, while the full lines
include the Breit-Wigner resonances. In the absence of an interference
term for the reaction to the $^8$Be ground state, the DWBA alone reproduces
the trend of the data points outside the resonance region. For the
reaction to the first excited state of $^8$Be, the DWBA curve is in
good overall agreement with the ad-hoc assumption of a direct, non-resonant
contribution made in Ref.~\cite{ang93}, with the exception of a decrease
of $S_{NR}$ with increasing energy.
Note the pronounced enhancement of the
low energy data points due to the effect of electron screening~\cite{ang93}.

Due to the strong contribution of interference effects to the $\alpha_1$
$S$-factor, one cannot hope to describe the angular distributions with
the DWBA. For the $\alpha_0$ $S$-factor, however, such a description
seems to be reasonable outside the resonance. In Ref.~\cite{ang93}
no differential cross sections were measured. In Ref.~\cite{beck87}
a few angular distribution curves are available, but they tend to
disagree with measurements presently carried out~\cite{sta93}. The
DWBA calculations favour Ref.~\cite{sta93}, but the exact shape of the
angular distributions depend crucially on the value $\lambda_p$ of
the proton-$^{11}$B folding potential.

The reaction rate $N_A \left< \sigma v \right>$ 
of $^{11}$B(p,$\alpha_1$)$^8$Be$^*$ is listed in Table~\ref{tab7} and compared
to the values obtained from the parametrization given in
Ref.~\cite{cau88} (here, the contribution of the reaction 
$^{11}$B(p,$\alpha_0$)$^8$Be is neglected since it is about $10^{-2}$
of the rate for the transition to the first excited state). 
There is a slight enhancement of the rate at low
temperatures due to the better description of the resonances and the
direct contribution.

\section{Conclusion}

The reactions $^{10}$B(p,$\alpha$)$^7$Be and $^{11}$B(p,$\alpha$)$^8$Be
are described well by the DWBA calculations. In the case of
$^{10}$B(p,$\alpha_0$)$^7$Be, we were able to reproduce the resonance
structure in a consistent way within our potential model thus suggesting
that it can be regarded as a potential resonance. The influence of the
resonance on the reaction rate can be seen very clearly in Fig.~\ref{fig3} (a
fairly constant $S$-factor was assumed in Ref.~\cite{cau88} at low
energies). This clearly demonstrates that
extrapolations from higher energies have to be done very carefully to
include the correct shape of the resonance.

For the transition to the first excited state in $^7$Be the
interference effects with a 5/2$^+$ level at about 550 keV have to be
taken into account. This was achieved by including a Breit-Wigner term
describing the resonance at 550 keV and an interference term between the
DWBA and Breit-Wigner contributions. However, the cross sections of the
$\alpha_1$ transition are lower
by several orders of magnitude than those of 
the $\alpha_0$ transition and therefore it
does not contribute significantly to the final reaction rate.

In the case of $^{11}$B(p,$\alpha$)$^8$Be the inclusion of an interference 
term
between single-level Breit-Wigner and DWBA
also reproduces the data acceptably well.
Systematic studies at higher
energies are being carried out at present \cite{sch94}.

\acknowledgements
We want to thank G.\ Staudt and H.\ Oberhummer for discussions and
comments on the manuscript, C.\ Angulo,
F.\ Knape, and F.-K. Thielemann for discussions, and H. Clement for 
pointing out the
results of pion scattering experiments. Our thanks also go 
to O. Sch\"onfeld who
supplied the $\alpha$-$^8$Be potential and to R.N. Boyd and M.J. Balbes 
for helpful comments on the manuscript and critical discussions.
GR wishes to express his thanks to C. Rolfs for support during his stay
in M\"unster and Bochum.
This research was supported in part by the
Deutsche Forschungsgemeinschaft (projects Ro429/18-2 and Ro429/21-3).
TR acknowledges support by the Alexander von Humboldt foundation and
by an APART fellowship of the Austrian Academy of Sciences. GR was
supported by an Ohio State University postdoctoral fellowship.

\newpage


\begin{figure}
\caption{$S$-factor data of the reaction
$^{10}$B(p,$\alpha_0$)$^7$Be in the range $E_{c.m.} \leq 0.04$ MeV. The
curve is the result of the DWBA calculation. The experimental data
are represented by triangles~\protect{\cite{ang93}}, and by 
squares~\protect{\cite{kna92b}}. }
\label{fig1a}
\end{figure}

\begin{figure}
\caption{$S$-factor data of the reaction
$^{10}$B(p,$\alpha_0$)$^7$Be in the range $0.04 \leq E_{c.m.} \leq 0.15$ MeV.
The curve is the result of the DWBA calculation. The experimental data
are represented by triangles~\protect{\cite{ang93}}, and by 
squares~\protect{\cite{kna92b}}. }
\label{fig1b}
\end{figure}

\begin{figure}
\caption{$S$-factor data of the reaction
$^{10}$B(p,$\alpha_1$)$^7$Be$^*$ in the energy range $E_{c.m.} \leq
0.2$ MeV. The
experimental data are taken from Ref.~\protect{\cite{ang93b}}.
Shown are the DWBA
contribution (dotted), the Breit-Wigner contribution (dashed) and the
sum of DWBA, Breit-Wigner and interference term (solid).}
\label{fig2}
\end{figure}

\begin{figure}
\caption{Ratio of the reaction rate of
$^{10}$B(p,$\alpha$)$^7$Be obtained in this work
to rate values from Refs.~\protect{\cite{cau75,cau88}}. }
\label{fig3}
\end{figure}

\begin{figure}
\caption{$S$-factor data of the reactions
$^{11}$B(p,$\alpha_1$)$^8$Be$^*$ and $^{11}$B(p,$\alpha_0$)$^8$Be.
Dashed curves: non-resonant contribution (DWBA); solid curves: contribution 
from the sum of DWBA, Breit-Wigner and interference terms.
The data are taken from 
Refs.~\protect{\cite{beck87}} (triangles) 
and \protect{\cite{ang93}} (circles). }
\label{fig4b}
\end{figure}

\newpage


\begin{table}
\caption{Spectroscopic factors for $^{10}$B = t + $^7$Be.}

\begin{tabular}{ccccc}
$J (^7$Be)  & $E_x$ ($^7$Be)\tablenotemark[1]&P$_{3/2}$&F$_{5/2}$ & F$_{7/2}$ 
\\
\tableline
1/2	    & \dec 1.1 MeV    &	  & \dec 0.0136    & \dec 0.0037  \\
3/2	    & \dec 0.0 MeV    & \dec 0.0812  & \dec 0.0706    & \dec 0.2571 
\\
\end{tabular}
\tablenotetext[1]{$E_x$ is the excitation energy calculated from the
shell model~\cite{kur75}.}
\label{tab1}
\end{table}


\begin{table}
\caption{Parameters of the optical and bound state potentials
for the reaction $^{10}$B(p,$\alpha$)$^{7}$Be.}
\begin{tabular}{ll}
p + $^{10}$B & Real part: single-folding potential \\
             & $\lambda _p$ = 1.326      \\
             & $r_c$ = 1.2 fm \tablenotemark[1]  \\
             & Imaginary part: Saxon-Woods derivative potential       \\
             & $W_D$ = $-15$ keV, $r_D$ = 1.5 fm\tablenotemark[1], $a_D$ = 
0.5
fm\tablenotemark[1] \\ \\
$\alpha$ + $^{7}$Be
             & Real part: double-folding potential \\
             & $\lambda _{\alpha}$ = 1.7      \\
             & $r_c$ = 1.69 fm \tablenotemark[1] \\
             & Imaginary part: Saxon-Woods volume potential       \\
             & $W_V$ = $-4.6$ MeV, $r_V$ = 1.4 fm \tablenotemark[1], 
$a_V$ = 0.52 fm \tablenotemark[1] \\ \\
Bound state  & double-folding potential \\
(t + $^7$Be) & $\lambda$ \tablenotemark[2] \\
             & $r_c$ = 1.4 fm \tablenotemark[1]                    \\
\end{tabular}
\tablenotetext[1]{Taken from Ref.\ \cite{per76}.}
\tablenotetext[2]{Calculated for the different separation energies
corresponding to the different states of $^{7}$Be~\cite{ajz88}
(see text for more information).}
\label{tab2}
\end{table}
\newpage

\begin{table}
\caption{Spectroscopic factors for $^{11}$B = t + $^8$Be.}

\begin{tabular}{cccccc}
$J (^8$Be)  &$E_x$ ($^8$Be)\tablenotemark[1] & 
		P$_{1/2}$&P$_{3/2}$&F$_{5/2}$&F$_{7/2}$\\
\tableline
0	    & \dec 0.0 MeV & & \dec 0.2632    &		  &	     \\
2& \dec 3.4 MeV & \dec 0.0001 & \dec 0.4669 & \dec 0.0618 & \dec 0.0577   \\
\end{tabular}
\tablenotetext[1]{$E_x$ is the excitation energy calculated from the
shell model~\cite{kur75}.}
\label{tab3}
\end{table}
\newpage
%

%
\begin{table}
\caption{Parameters of the optical and bound state potentials
for the reaction $^{11}$B(p,$\alpha$)$^{8}$Be.}
\begin{tabular}{ll}
p + $^{11}$B & Real part: single-folding potential \\
             & $\lambda _p$ = 0.81                 \\
             & $r_c$ = 1.29 fm\tablenotemark[1]  \\ \\
$\alpha$ + $^{8}$Be
             & Real part: double-folding potential \\
             & $\lambda _{\alpha}$ = 1.21          \\
             & $r_c$ = 1.55 fm \tablenotemark[1]\tablenotemark[2] \\
             & Imaginary part: Saxon-Woods volume potential       \\
             & $W_V$ = $-3$ MeV, $r_V$ = 1.75 
fm\tablenotemark[1]\tablenotemark[2],
$a_V$ = 0.65 fm\tablenotemark[1]\tablenotemark[2] \\ \\
Bound state  & double-folding potential \\
(t + $^8$Be) & $\lambda$ \tablenotemark[3] \\
             & $r_c$ = 1.5 fm\tablenotemark[1]\tablenotemark[2]      \\
\end{tabular}
\tablenotetext[1]{Taken from Ref.\ \cite{per76}.}
\tablenotetext[2]{Averaged (see text).}
\tablenotetext[3]{Calculated for the different separation energies
corresponding to the different states of $^{8}$Be~\cite{ajz88} (see text
for more information).}
\label{tab4}
\end{table}
\newpage


\begin{table}
\caption{Results of the Breit-Wigner fits for the reaction
$^{11}$B(p,$\alpha$)$^8$Be.}

\begin{tabular}{lll}
Reaction &	 Resonance 1		    & Resonance 2		   \\
\tableline
$^{11}$B(p,$\alpha_0$)$^8$Be
&		 $\theta_p^2    = 0.017$   & \\	
&		 $\Gamma_\alpha = 5.5$ keV & \\
&		 \\
&		 \\
$^{11}$B(p,$\alpha_1$)$^8$Be$^*$ 
& $\theta_p^2    = 0.570$   & $\theta_p^2    = 0.604$      \\
&		 $\Gamma_\alpha = 5.7$ keV & $\Gamma_\alpha = 296.5$ keV  \\
&		 \\
&		 $E_{c.m.} = 148.5$ keV\tablenotemark[1] 
					    & $E_{c.m.} = 660$ keV    
\end{tabular}
\tablenotetext[1]{No fit parameter; value was taken from Ref.\
\cite{ajz90}.}
\label{tab5}
\end{table}
\newpage

\begin{table}
\caption{Reaction rate $N_A \left< \sigma v \right>$ of the
reaction $^{10}$B(p,$\alpha$)$^7$Be in cm$^3$s$^{-1}$mole$^{-1}$. The
rate calculated with DWBA for bare nuclei is compared to values given in
previous work.}

\begin{tabular}{lr@{$\times 10$}lr@{$\times 10$}lr@{$\times 10$}l}
Temperature \tablenotemark[1] & 
\multicolumn{2}{c}{Caughlan {\it et al.} \tablenotemark[2]} &
\multicolumn{2}{c}{Youn {\it et al.} \tablenotemark[3]} &
\multicolumn{2}{c}{this work} \\
\tableline
0.002 & 0.209&$^{-28}$ &\multicolumn{2}{l}{ }&	0.494&$^{-26}$ \\
0.004 & 0.496&$^{-20}$ &\multicolumn{2}{l}{ }&	0.175&$^{-17}$ \\
0.006 & 0.561&$^{-16}$ &\multicolumn{2}{l}{ }&	0.208&$^{-13}$ \\
0.008 & 0.200&$^{-13}$ &\multicolumn{2}{l}{ }&	0.630&$^{-11}$ \\
0.010 & 0.129&$^{-11}$ & 0.167&$^{-9}$ &	0.323&$^{-9}$ \\
0.012 & 0.310&$^{-10}$ & 0.319&$^{-8}$ &	0.607&$^{-8}$ \\
0.014 & 0.391&$^{-9}$ & 0.326&$^{-7}$ &	0.605&$^{-7}$ \\
0.016 & 0.315&$^{-8}$ & 0.217&$^{-6}$ &	0.393&$^{-6}$ \\
0.018 & 0.184&$^{-7}$ & 0.107&$^{-5}$ &	0.188&$^{-5}$ \\
0.020 & 0.836&$^{-7}$ & 0.416&$^{-5}$ &	0.716&$^{-5}$ \\
0.025 & 0.173&$^{-5}$ & 0.617&$^{-4}$ &	0.102&$^{-3}$ \\
0.030 & 0.174&$^{-4}$ & 0.471&$^{-3}$ &	0.754&$^{-3}$ \\
0.035 & 0.109&$^{-3}$ &\multicolumn{2}{l}{ }&	0.369&$^{-2}$ \\
0.040 & 0.495&$^{-3}$ & 0.880&$^{-2}$ &	0.136&$^{-1}$ \\
0.045 & 0.177&$^{-2}$ &\multicolumn{2}{l}{ }&	0.406&$^{-1}$ \\
0.050 & 0.530&$^{-2}$ & 0.687&$^{-1}$ &	0.104&$^{0}$       \\
0.060 & 0.321&$^{-1}$ & 0.324&$^{0}$       &	0.484&$^{0}$       \\
0.070 & 0.135&$^{0}$       & 0.110&$^{1}$   &	0.164&$^{1}$   \\
0.080 & 0.437&$^{0}$       & 0.301&$^{1}$   &	0.448&$^{1}$ \\
0.090 & 0.118&$^{1}$   & 0.700&$^{1}$   &	0.104&$^{2}$ \\
0.100 & 0.277&$^{1}$   & 0.144&$^{2}$   &	0.214&$^{2}$ \\
0.120 & 0.112&$^{2}$   & 0.468&$^{2}$   &	0.700&$^{2}$ \\
0.140 & 0.341&$^{2}$   & 0.119&$^{3}$   &	0.179&$^{3}$ \\
0.160 & 0.850&$^{2}$   & 0.254&$^{3}$   &	0.387&$^{3}$ \\
0.180 & 0.183&$^{3}$   & 0.481&$^{3}$   &	0.739&$^{3}$ \\
0.200 & 0.354&$^{3}$   & 0.831&$^{3}$   &	0.129&$^{4}$ \\
0.300 & 0.355&$^{4}$   & 0.552&$^{4}$   &	0.899&$^{4}$ \\
0.400 & 0.149&$^{5}$   & 0.179&$^{5}$   &	0.302&$^{5}$ \\
0.500 & 0.412&$^{5}$   & 0.415&$^{5}$   &	0.707&$^{5}$ \\
0.600 & 0.891&$^{5}$   & 0.814&$^{5}$   &	0.134&$^{6}$ \\
0.700 & 0.165&$^{6}$   & 0.145&$^{6}$   &	0.219&$^{6}$ \\
0.800 & 0.276&$^{6}$   & 0.240&$^{6}$   &	0.324&$^{6}$ \\
0.900 & 0.430&$^{6}$   & 0.379&$^{6}$   &	0.442&$^{6}$ \\
1.000 & 0.634&$^{6}$   & 0.571&$^{6}$   &	0.570&$^{6}$ \\
\end{tabular}
\tablenotetext[1]{Given in 10$^9$ K.}
\tablenotetext[2]{Calculated with the parametrization given in
Refs.\ \cite{cau75,cau88}.}
\tablenotetext[3]{Here we cite Ref.\ \cite{you91}.}
\label{tab6}
\end{table}
\newpage

\begin{table}
\caption{Reaction rates $N_A \left< \sigma v \right>$ of the
reaction $^{11}$B(p,$\alpha$)$^8$Be$^*$ in cm$^3$s$^{-1}$mole$^{-1}$. The
rate calculated with DWBA is compared to values given in previous
work.}

\begin{tabular}{lr@{$\times 10$}lr@{$\times 10$}l}
Temperature \tablenotemark[1] & 
\multicolumn{2}{c}{Caughlan {\it et al.} \tablenotemark[2]} &
\multicolumn{2}{c}{this work} \\
\tableline
 0.002   &   0.284&$^{-27}$   &   0.197&$^{-31}$   \\
 0.004   &   0.715&$^{-19}$   &   0.281&$^{-19}$   \\
 0.006   &   0.834&$^{-15}$   &   0.613&$^{-15}$   \\
 0.008   &   0.303&$^{-12}$   &   0.383&$^{-12}$   \\
 0.010   &   0.200&$^{-10}$   &   0.255&$^{-10}$   \\
 0.012   &   0.486&$^{-9}$   &   0.563&$^{-9}$   \\
 0.014   &   0.618&$^{-8}$   &   0.702&$^{-8}$   \\
 0.016   &   0.504&$^{-7}$   &   0.582&$^{-7}$   \\
 0.018   &   0.296&$^{-6}$   &   0.349&$^{-6}$   \\
 0.020   &   0.136&$^{-5}$   &   0.162&$^{-5}$   \\
 0.025   &   0.286&$^{-4}$   &   0.345&$^{-4}$   \\
 0.030   &   0.291&$^{-3}$   &   0.353&$^{-3}$   \\
 0.035   &   0.185&$^{-2}$   &   0.226&$^{-2}$   \\
 0.040   &   0.849&$^{-2}$   &   0.104&$^{-1}$   \\
 0.045   &   0.307&$^{-1}$   &   0.377&$^{-1}$   \\
 0.050   &   0.927&$^{-1}$   &   0.114&$^{0}$   \\
 0.060   &   0.571&$^{0}$   &   0.708&$^{0}$   \\
 0.070   &   0.244&$^{1}$   &   0.304&$^{1}$   \\
 0.080   &   0.808&$^{1}$   &   0.102&$^{2}$   \\
 0.090   &   0.226&$^{2}$   &   0.286&$^{2}$   \\
 0.100   &   0.558&$^{2}$   &   0.711&$^{2}$   \\
 0.120   &   0.267&$^{3}$   &   0.339&$^{3}$   \\
 0.140   &   0.990&$^{3}$   &   0.123&$^{4}$   \\
 0.160   &   0.293&$^{4}$   &   0.358&$^{4}$   \\
 0.180   &   0.717&$^{4}$   &   0.864&$^{4}$   \\
 0.200   &   0.150&$^{5}$   &   0.180&$^{5}$   \\
 0.300   &   0.153&$^{6}$   &   0.187&$^{6}$   \\
 0.400   &   0.552&$^{6}$   &   0.711&$^{6}$   \\
 0.500   &   0.138&$^{7}$   &   0.183&$^{7}$   \\
 0.600   &   0.298&$^{7}$   &   0.381&$^{7}$   \\
 0.700   &   0.588&$^{7}$   &   0.689&$^{7}$   \\
 0.800   &   0.106&$^{8}$   &   0.112&$^{8}$   \\
 0.900   &   0.174&$^{8}$   &   0.167&$^{8}$   \\
 1.000   &   0.264&$^{8}$   &   0.233&$^{8}$   \\
\end{tabular}
\tablenotetext[1]{Given in 10$^9$ K.}
\tablenotetext[2]{Here we cite Ref.\ \cite{cau88}.}
\label{tab7}
\end{table}

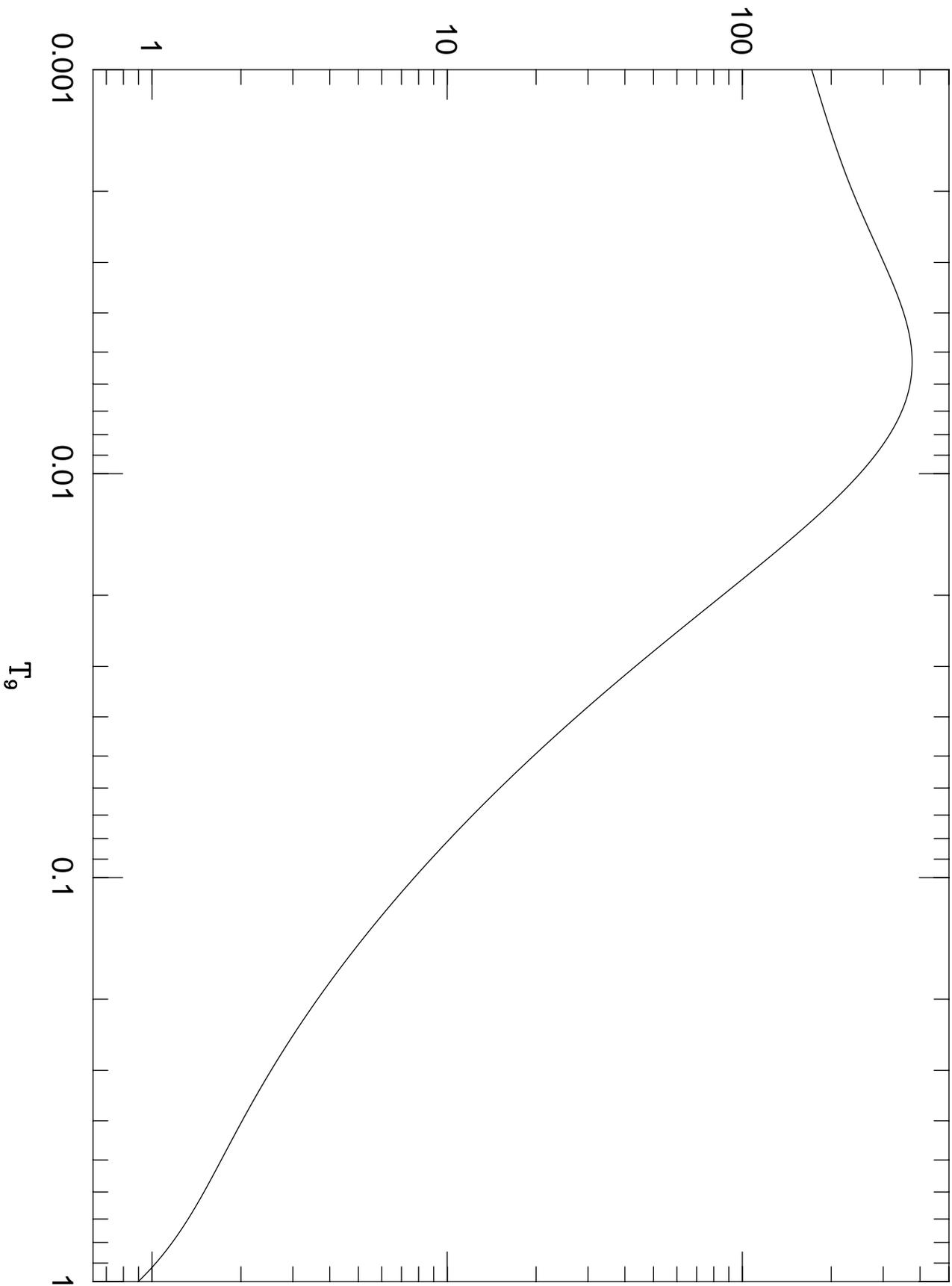
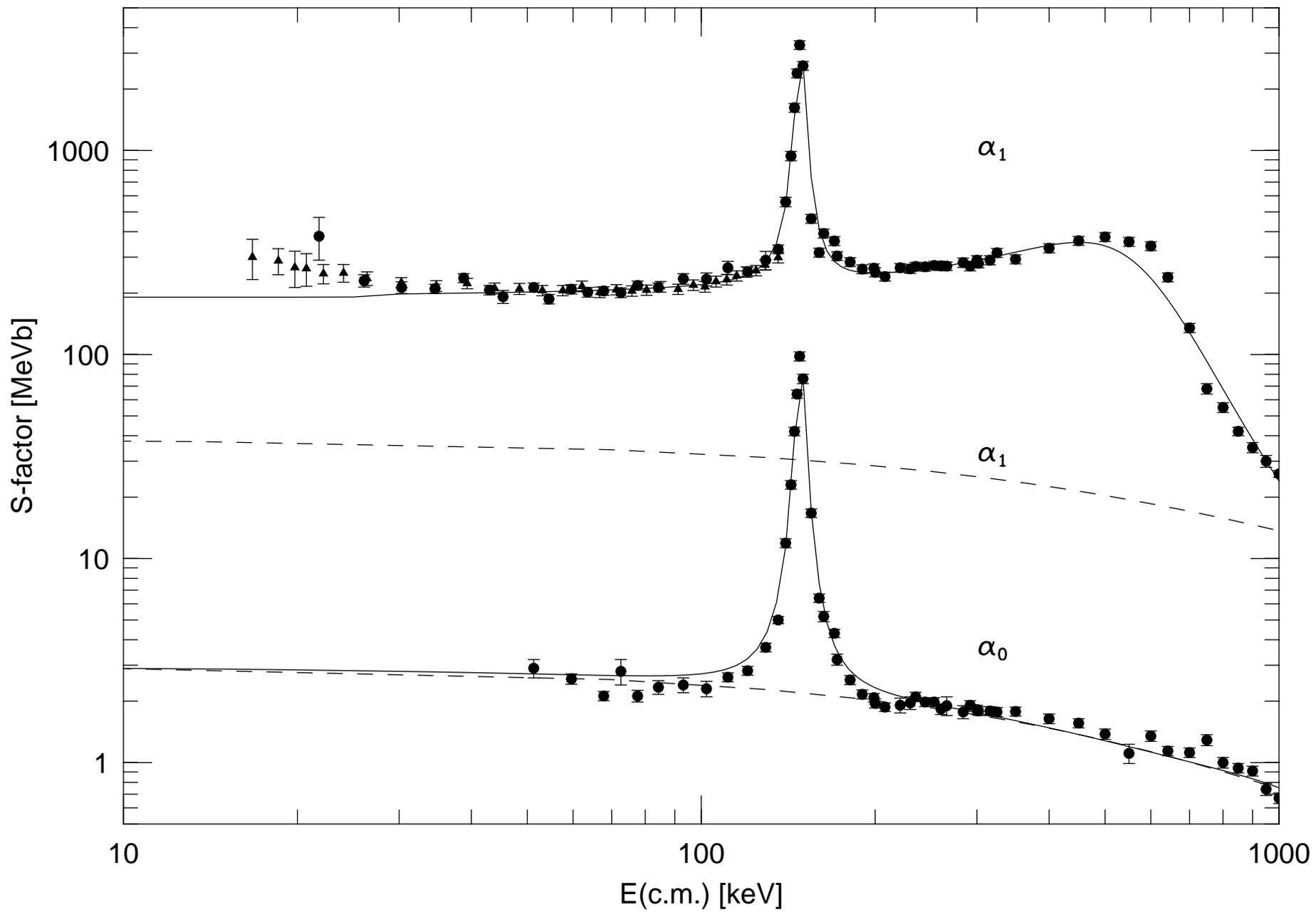
\end{document}